\documentclass[twoside,twocolumn,9pt]{article}
\usepackage{extsizes}
\usepackage[super,sort&compress,comma]{natbib} 
\usepackage[version=3]{mhchem}
\usepackage[left=1.5cm, right=1.5cm, top=1.785cm, bottom=2.0cm]{geometry}
\usepackage{balance}
\usepackage{mathptmx}
\usepackage{sectsty}
\usepackage{graphicx} 
\usepackage{lastpage}
\usepackage[format=plain,justification=justified,singlelinecheck=false,font={stretch=1.125,small,sf},labelfont=bf,labelsep=space]{caption}
\usepackage{float}
\usepackage{comment}
\usepackage{fancyhdr}
\usepackage{fnpos}
\usepackage[english]{babel}
\addto{\captionsenglish}{%
  
}
\usepackage{array}
\usepackage{droidsans}
\usepackage{charter}
\usepackage[T1]{fontenc}
\usepackage[usenames,dvipsnames]{xcolor}
\usepackage{setspace}
\usepackage[compact]{titlesec}
\usepackage{hyperref}
\usepackage[normalem]{ulem}
\usepackage{enumitem}
\usepackage{epstopdf}%This line makes .eps figures into .pdf - please comment out if not required.

\usepackage{gensymb}
\usepackage{multirow}

\definecolor{cream}{RGB}{222,217,201}

\begin{document}

\pagestyle{fancy}
\thispagestyle{plain}
\fancypagestyle{plain}{
%%%HEADER%%%
\renewcommand{\headrulewidth}{0pt}
}
%%%END OF HEADER%%%

%%%PAGE SETUP - Please do not change any commands within this section%%%
\makeFNbottom
\makeatletter
\renewcommand\LARGE{\@setfontsize\LARGE{15pt}{17}}
\renewcommand\Large{\@setfontsize\Large{12pt}{14}}
\renewcommand\large{\@setfontsize\large{10pt}{12}}
\renewcommand\footnotesize{\@setfontsize\footnotesize{7pt}{10}}
\makeatother

\renewcommand{\thefootnote}{\fnsymbol{footnote}}
\renewcommand\footnoterule{\vspace*{1pt}% 
\color{cream}\hrule width 3.5in height 0.4pt \color{black}\vspace*{5pt}} 
\setcounter{secnumdepth}{5}

\makeatletter 
\renewcommand\@biblabel[1]{#1}            
\renewcommand\@makefntext[1]% 
{\noindent\makebox[0pt][r]{\@thefnmark\,}#1}
\makeatother 
\renewcommand{\figurename}{\small{Fig.}~}
\sectionfont{\sffamily\Large}
\subsectionfont{\normalsize}
\subsubsectionfont{\bf}
\setstretch{1.125} %In particular, please do not alter this line.
\setlength{\skip\footins}{0.8cm}
\setlength{\footnotesep}{0.25cm}
\setlength{\jot}{10pt}
\titlespacing*{\section}{0pt}{4pt}{4pt}
\titlespacing*{\subsection}{0pt}{15pt}{1pt}
%%%END OF PAGE SETUP%%%

%%%FOOTER%%%
\fancyfoot{}
\fancyfoot[RO]{\footnotesize{\sffamily{  \hspace{2pt}\thepage}}}
\fancyfoot[LE]{\footnotesize{\sffamily{\thepage}}}
\fancyhead{}
\renewcommand{\headrulewidth}{0pt} 
\renewcommand{\footrulewidth}{0pt}
\setlength{\arrayrulewidth}{1pt}
\setlength{\columnsep}{6.5mm}
\setlength\bibsep{1pt}
%%%END OF FOOTER%%%

%%%FIGURE SETUP - please do not change any commands within this section%%%
\makeatletter 
\newlength{\figrulesep} 
\setlength{\figrulesep}{0.5\textfloatsep} 

\newcommand{\topfigrule}{\vspace*{-1pt}% 
\noindent{\color{cream}\rule[-\figrulesep]{\columnwidth}{1.5pt}} }

\newcommand{\botfigrule}{\vspace*{-2pt}% 
\noindent{\color{cream}\rule[\figrulesep]{\columnwidth}{1.5pt}} }

\newcommand{\dblfigrule}{\vspace*{-1pt}% 
\noindent{\color{cream}\rule[-\figrulesep]{\textwidth}{1.5pt}} }

\makeatother
%%%END OF FIGURE SETUP%%%

%%%TITLE, AUTHORS AND ABSTRACT%%%
\twocolumn[
  \begin{@twocolumnfalse}
%{\includegraphics[height=30pt]{head_foot/SM}\hfill\raisebox{0pt}[0pt][0pt]{\includegraphics[height=55pt]{head_foot/RSC_LOGO_CMYK}}\\[1ex]
%\includegraphics[width=18.5cm]{head_foot/header_bar}}\par
\vspace{1em}
\sffamily
\begin{tabular}{m{4.5cm} p{13.5cm} }

 & \noindent\LARGE{\textbf{Active flows drive anchoring of nematics at rigid walls}} \\%Article title goes here instead of the text "This is the title"
\vspace{0.3cm} & \vspace{0.3cm} \\

 % & \noindent\large{Full Name,$^{\ast}$\textit{$^{a}$} Full Name,\textit{$^{b\ddag}$} and Full Name\textit{$^{a}$}} \\%Author names go here instead of "Full name", etc.
  & \noindent\large{
  Michael Fang,\textit{$^{a,b}$} 
  Ioannis Hadjifrangiskou,\textit{$^{b}$} 
  Sumesh P. Thampi,\textit{$^{c}$}
  Julia M. Yeomans\textit{$^{b}$}
  and Jan Rozman,$^{\ast}$\textit{$^{d,e}$}} \\ \\%Author names go here instead of "Full name", etc.

 & 
{Although confinement strongly influences flows in active materials, it remains unclear how active particles align at rigid boundaries when no thermodynamic anchoring is imposed. We address this question using continuum simulations of active nematics, together with analytical arguments based on a reduced near-wall description. In the flow-tumbling regime, extensile systems align parallel to the boundary, whereas contractile systems align perpendicular to it, consistent with active anchoring observed at active-passive interfaces. In the flow-aligning regime, the preferred orientation depends on the sign of activity and of the flow aligning parameter: either the shear-like flow generated near the wall selects the Leslie angle, or no unique alignment is established. These results provide a unified framework for activity-induced anchoring at rigid walls, demonstrating that boundary alignment in dense active matter can emerge solely from the interplay between self-generated flows and orientational dynamics.} \\

\end{tabular}

 \end{@twocolumnfalse} \vspace{0.6cm}

  ]
%%%END OF TITLE, AUTHORS AND ABSTRACT%%%

%%%FONT SETUP - please do not change any commands within this section
\renewcommand*\rmdefault{bch}\normalfont\upshape
\rmfamily
\section*{}
\vspace{-1cm}

%%%FOOTNOTES%%%

\footnotetext{\textit{$^{a}$~Department of Physics, Princeton University, Princeton, NJ 08544, USA.}}
\footnotetext{\textit{$^{b}$~Rudolf Peierls Centre for Theoretical Physics, University of Oxford, Oxford OX1 3PU, United Kingdom}}
\footnotetext{\textit{$^{c}$~Department of Chemical Engineering, Indian Institute of Technology, Madras, Chennai, India 600036.}}
\footnotetext{\textit{$^{d}$~Jo\v zef Stefan Institute, Jamova 39, SI-1000 Ljubljana, Slovenia; E-mail: jan.rozman@ijs.si}}
\footnotetext{\textit{$^{e}$~University of Ljubljana, Faculty of Mathematics and Physics, Jadranska 19, SI-1000 Ljubljana, Slovenia.}}
%Please use \dag to cite the ESI in the main text of the article.
%If you article does not have ESI please remove the the \dag symbol from the title and the footnotetext below.
%\footnotetext{\dag~Supplementary Information available: [details of any supplementary information available should be included here]. See DOI: 10.1039/cXsm00000x/}
%additional addresses can be cited as above using the lower-case letters, c, d, e... If all authors are from the same address, no letter is required

% \footnotetext{\dag~Additional footnotes to the title and authors can be included \textit{e.g.}\ `Present address:' or `These authors contributed equally to this work' as above using the symbols: \ddag, \textsection, and \P. Please place the appropriate symbol next to the author's name and include a \texttt{\textbackslash footnotetext} entry in the the correct place in the list.}

%\ddag for both dagguers amd reintroduce the part about the SI
%%%END OF FOOTNOTES%%%

%%%MAIN TEXT%%%%

%||||||||||||||||||||||||||||||||||||||||||||||||||||||||||||
%||||||||||||||||||||||||||||||||||||||||||||||||||||||||||||
%||||||||||||||||||||||INTRODUCTION||||||||||||||||||||||||||
%||||||||||||||||||||||||||||||||||||||||||||||||||||||||||||
%||||||||||||||||||||||||||||||||||||||||||||||||||||||||||||

\section{Introduction}

A distinctive feature of many wet active systems, including suspensions of bacteria and microtubule-motor mixtures, is active turbulence, a state of chaotic flow characterised by velocity jets, vortices, and the continual creation and annihilation of topological defects.\cite{alert2022active, doostmohammadi2018active, martinez2019selection} Active turbulence is well described by continuum theories of active nematics, in which the equations of nematic liquid-crystal hydrodynamics are supplemented by an active stress. Under confinement, the instabilities responsible for active turbulence can be suppressed, leading instead to coherent laminar flows or regular vortex states.\cite{voituriez2005spontaneous, thampi2022channel, hardouin2022active, shendruk2017dancing, opathalage2019self-organised} Understanding how active nematics orient at confining boundaries is therefore important in controlling active flows near surfaces and in microfluidic environments.\cite{guillamat2017taming, pearce2019geometrical, vcopar2019topology, varghese2020confinement, wu2017transition, lushi2014fluid, wioland2013confinement, duclos2018spontaneous, dong2024collective, joshi2023disks, peng2016command}

In passive liquid crystals, particles can align parallel to confining surfaces to maximize their translational entropy.\cite{garlea2019colloidal,dijkstra2001wetting} Molecular alignment at surfaces can also be externally imposed through specific chemical treatments, rubbing techniques, or patterned substrates.\cite{de1993physics} At the continuum level, such anchoring is commonly modelled by adding a suitable surface contribution to the free energy, and similar terms are employed in theories and simulations of active nematics.\cite{fournier2005modeling} 

Active systems can, however, exhibit boundary alignment even in the absence of imposed anchoring.\cite{bhattacharyya2025active} A striking example is active anchoring at interfaces between active and passive fluids, where gradients in the active stress generate flows that orient particles relative to the interface.\cite{blow2014biphasic,coelho2021director}  In the context of individual bacteria it is well established that contractile microswimmers, which draw fluid inwards along their swimming axis, tend to align perpendicular to boundaries, whereas extensile swimmers, which push fluid outwards along their swimming axis preferentially orient parallel.\cite{kurzthaler2023hydrodynamics, berke2008hydrodynamic,spagnolie2012hydrodynamics} While activity-induced alignment is well established at active-passive interfaces and for individual microswimmers near surfaces, how similar mechanisms govern the orientation of dense active nematics at rigid walls remains unknown.

The orientational response of nematics to flow is known to depend sensitively on the flow aligning parameter.\cite{doi1988theory}  Flow-aligning systems tend to adopt a stable angle relative to an imposed shear flow, known as the Leslie angle, whereas flow-tumbling systems undergo continual rotation. The interplay between activity-induced flows and flow-induced alignment is expected to play a central role in determining the preferred orientation at the walls. 

In this paper we investigate how activity-induced flows determine the orientation of active nematics at rigid walls in the absence of imposed anchoring  by considering a 2D active fluid in a wide channel with turbulent flows in the bulk. We combine numerical solutions of the active nematic equations of motion with analytical arguments, to characterize the range of boundary alignment states adopted. In the flow-tumbling regime, activity generates robust anchoring analogous to that observed at active-passive interfaces, with extensile systems aligning parallel to walls and contractile systems aligning perpendicular to them. In contrast, in the flow-aligning regime the wall-generated flow either selects the Leslie angle or fails to produce a unique preferred orientation, depending on the type of activity and the sign of the flow aligning parameter.

The paper is organized as follows: In Section~\ref{sec:model}, we describe the active nematic model and simulation methods. Section~\ref{sec:results} presents our main results. Specifically, Section~\ref{subsec:tumbling} describes wall alignment of the director field in simulations of active nematics, showing that the behaviour differs between flow-aligning and flow-tumbling materials. Section~\ref{subsec:analytics} provides an analytical explanation for the observed alignments, and the analytical predictions are compared to the simulations in Section~\ref{subsec:comparisons}. Finally, Section~\ref{sec:conclusions} summarizes our findings and concludes the paper.

%||||||||||||||||||||||||||||||||||||||||||||||||||||||||||||
%||||||||||||||||||||||||||||||||||||||||||||||||||||||||||||
%||||||||||||||||||||||MODEL|||||||||||||||||||||||||||||||||
%||||||||||||||||||||||||||||||||||||||||||||||||||||||||||||
%||||||||||||||||||||||||||||||||||||||||||||||||||||||||||||
\section{Model}\label{sec:model}

We model the active fluid by solving the nematohydrodynamic equations of motion. A second rank traceless, symmetric tensor, the nematic order parameter $\mathbf{Q}$, defines the microstructure of the fluid. $\mathbf{Q}$  evolves according to\cite{simha2002hydrodynamic, doostmohammadi2018active} 
\begin{align}
\partial_t \mathbf{Q} + \mathbf{u}\cdot \boldsymbol{\nabla} \mathbf{Q} - \mathbf{S}_{\mathbf{Q}} = \Gamma_{\mathbf{Q}} \mathbf{H}_{\mathbf{Q}}.\label{qdynamics}
\end{align} 
Here $\Gamma_{\mathbf{Q}}$ determines the relaxation time, and $\mathbf{H}_{\mathbf{Q}}$ is the molecular potential, calculated from the variational derivative of the free energy $\mathcal{F}_{\mathbf{Q}}$,
\begin{align}
    \mathbf{H}_{\mathbf{Q}} = -\frac{\delta\mathcal{F}_{\mathbf{Q}}}{\delta\mathbf{Q}} + \frac{\mathbf{I}}{3} \textnormal{Tr}\left(\frac{\delta\mathcal{F}_{\mathbf{Q}}}{\delta\mathbf{Q}}\right).
\end{align}
The free energy density in $\mathcal{F}_{\mathbf{Q}} = \int f_{\mathbf{Q}}\mathrm{d}V$ follows the Landau-de Gennes prescription:
\begin{align}
    f_{\mathbf{Q}} = \frac{A_0}{2}\left(1-\frac{\gamma}{3}\right)Q_{ij}^2 &- \frac{1}{3} A_0 \gamma Q_{ij}Q_{jk}Q_{ki} + \frac{1}{4} A_0\gamma (Q_{ij}^2)^2 \nonumber\\
    &+ \frac{K_{\mathbf{Q}}}{2} \left(\partial_k Q_{ij}\right)^2,
    \label{eq:FQ}
\end{align}
where $A_0$ is a phenomenological constant and $\gamma$ specifies the temperature. We consider $\gamma > 3$, corresponding to the system being in the nematic state in the absence of activity.\cite{chandragiri2019active} We use a single elastic constant approximation with elastic constant $K_{\mathbf{Q}}$.

In Eq.~\eqref{qdynamics}, $\mathbf{S}_{\mathbf{Q}}$ describes the response of the order parameter to velocity gradients,
\begin{align}
    \mathbf{S}_{\mathbf{Q}} &= \left(\xi\mathbf{E}+\boldsymbol{\Omega}\right)\cdot\left(\mathbf{Q}+\frac{\mathbf{I}}{3}\right) + \left(\mathbf{Q}+\frac{\mathbf{I}}{3}\right) \cdot \left(\xi\mathbf{E}-\boldsymbol{\Omega}\right)\nonumber\\
    &-2\xi \left(\mathbf{Q}+\frac{\mathbf{I}}{3}\right) \left(\mathbf{Q}:\nabla\mathbf{u}\right),
\end{align}
where $\mathbf{E}$ and $\boldsymbol{\Omega}$ respectively represent the symmetric and antisymmetric part of the velocity gradient tensor and $\xi$ is the flow aligning parameter. 
 
The velocity field $\mathbf{u}$ follows the incompressible Navier-Stokes equations
\begin{align}
\nabla\cdot\mathbf{u} &=0, \label{eqn:continuity}\\
\rho \left(\partial_t + \mathbf{u}\cdot\nabla\right) \mathbf{u} &= \nabla\cdot\boldsymbol{\Pi},\label{eqn:ns}
\end{align}
where $\rho$ is the density. The total stress tensor $\boldsymbol{\Pi}$ includes three contributions:
\begin{itemize}
    \item the viscous stress
    \begin{align}
        \boldsymbol{\Pi}^\mathrm{viscous} = 2\eta\mathbf{E},
    \end{align}
    where $\eta$ is the viscosity of the active nematic fluid.  
    \item the elastic stress
    \begin{align}
       \boldsymbol{\Pi}^\mathrm{elastic} &= -P \mathbf{I} + 2\xi \left(\mathbf{Q} + \frac{\mathbf{I}}{3} \right)(\mathbf{Q}:\mathbf{H}_{\mathbf{Q}}) \nonumber\\ &- \xi\mathbf{H}_{\mathbf{Q}}\cdot\left(\mathbf{Q} + \frac{\mathbf{I}}{3} \right) - \xi\left(\mathbf{Q} + \frac{\mathbf{I}}{3} \right)\cdot \mathbf{H}_{\mathbf{Q}} \nonumber\\ &- \nabla\mathbf{Q}:\frac{\delta\mathcal{F}_{\mathbf{Q}}}{\delta\nabla\mathbf{Q}} + \mathbf{Q}\cdot\mathbf{H}_{\mathbf{Q}} - \mathbf{H}_{\mathbf{Q}} \cdot \mathbf{Q},
    \end{align}
     where $P$ is the pressure field.
    \item the active stress~\cite{simha2002hydrodynamic}
    \begin{align}
       \boldsymbol{\Pi}^\mathrm{active} = -\zeta \mathbf{Q},
    \end{align}
    where the proportionality constant $\zeta$ gives the activity strength. Extensile active nematics have $\zeta > 0$; contractile active nematics have $\zeta < 0$.\cite{doostmohammadi2018active, doostmohammadi2021physics}
\end{itemize}

The governing equations are solved numerically using a hybrid lattice Boltzmann method.\cite{thampi2014vorticity} In this framework the fluid dynamics is computed with a lattice Boltzmann scheme based on a $D3Q19$ velocity set. The evolution of the nematic order parameter $\mathbf{Q}$ is treated separately by solving the associated convection–diffusion equation using a method-of-lines formulation, where spatial derivatives are approximated with second-order central finite differences and time integration is carried out using an explicit Euler step. Although the lattice Boltzmann algorithm used employs a three-dimensional grid, the system is restricted to two dimensions by applying periodic boundary conditions along the third direction. Periodic boundaries are also imposed along open sides of the domain. For solid boundaries, no-slip conditions are enforced for the velocity field, while free-anchoring conditions are applied to the nematic order parameter. All simulations are continued until a statistically steady state is reached, after which measurements are collected. Parameter values which are used, unless otherwise specified, are given in Table~\ref{table:parameters}.

\begin{table}[h]
\small
  \caption{\ Simulation parameters}
  \label{table:parameters}
  \begin{tabular*}{0.48\textwidth}{@{\extracolsep{\fill}}lll}
    \hline
    Parameter & Symbol & Value \\
    \hline
    Activity & $\zeta$ & $\pm 0.002$ \\
    Nematic elasticity & $K_\textbf{Q}$ & $0.001$ \\
    Flow aligning parameter & $\xi$ & $[-2,2]$ \\
    Viscosity & $\eta$ &0.83 \\
    Bulk free energy parameter & $A_0$ & 0.084 \\
    Temperature & $\gamma$ & 3.09 \\
    Molecular potential relaxation timescale & $\Gamma$ & {0.3} \\
    \hline
  \end{tabular*} 
\end{table}

%||||||||||||||||||||||||||||||||||||||||||||||||||||||||||||
%||||||||||||||||||||||||||||||||||||||||||||||||||||||||||||
%||||||||||||||||||||||||||||||||||||||||||||||||||||||||||||
%||||||||||||||||||||||RESULTS|||||||||||||||||||||||||||||||
%||||||||||||||||||||||||||||||||||||||||||||||||||||||||||||
%||||||||||||||||||||||||||||||||||||||||||||||||||||||||||||
%||||||||||||||||||||||||||||||||||||||||||||||||||||||||||||
\section{Results}\label{sec:results}

%||||||||||||||||||||||||||||||||||||||||||||||||||||||||||||
%||||||||||||||||||||||||||||||||||||||||||||||||||||||||||||
%||||||||||||||||||||||Alignment at walls||||||||||||||||||||
%||||||||||||||||||||||||||||||||||||||||||||||||||||||||||||
%||||||||||||||||||||||||||||||||||||||||||||||||||||||||||||

\subsection{Alignment at walls}\label{subsec:tumbling}
We solve the equations of motion for an active nematic confined within a wide channel of size $N_x = N_y = 256$, with walls oriented along the $x$ direction. Simulations are performed for both extensile ($\zeta > 0$) and contractile ($\zeta < 0$) systems over a range of values of the flow aligning parameter $\xi$. The system develops a state of active turbulence constituted by active flows, short-range nematic order and the continual creation and destruction of topological defects of charge $\pm \tfrac{1}{2}$.\cite{thampi2014vorticity, giomi2015geometry} 

\begin{figure}[h!]
\centering
  \includegraphics[height=16cm]{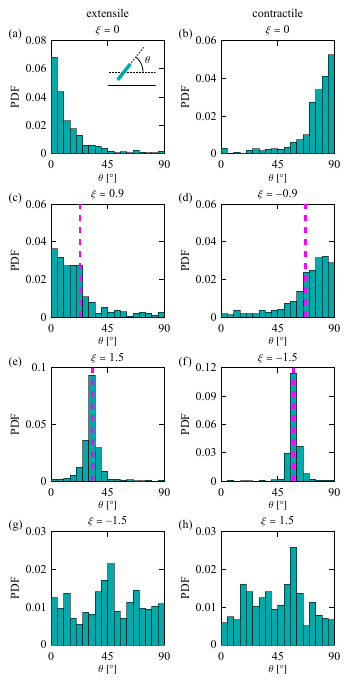}
  \caption{Distribution of angles between the director and the wall at $\xi=0$ (a,b), $\xi={\pm 0.9}$ (c,d), $\xi={\pm 1.5}$ (e,f), and $\xi={\mp 1.5}$ (g,h) for an extensile (a,c,e,g) and a contractile (b,d,f,h) system. Inset of panel (a) illustrates the plotted angle $\theta$, the smallest angle between the director and the wall. Dashed lines on (c)-(f) show the Leslie angle of the corresponding passive system in a simple shear flow [Eq.~\eqref{eq:leslie}].}
  \label{fig:distributions}
\end{figure}

We first consider the case where the flow aligning parameter $\xi = 0$, corresponding to the flow-tumbling regime. In this case, extensile active nematics exhibit predominantly planar alignment, with the director field oriented parallel to the wall, whereas contractile systems show primarily homeotropic alignment, with the director field oriented perpendicular to the wall. To quantify this behaviour, we define a director alignment angle $\theta$ at the wall as the smallest angle between the wall and the local director (inset of Fig.~\ref{fig:distributions}(a)). The magnitude of the alignment is characterized by plotting the probability distribution of $\theta$ for both systems, as shown in Fig.~\ref{fig:distributions}(a),(b). These distributions clearly demonstrate planar alignment in the extensile and homeotropic alignment in the contractile system.

%||||||||||||||||||||||||||||||||||||||||||||||||||||||||||||
%||||||||||||||||||||||FIGURE 1||||||||||||||||||||||||||||||
%||||||||||||||||||||||||||||||||||||||||||||||||||||||||||||

In the numerical simulations, the wall boundary condition on the $\mathbf{Q}$ field is free anchoring, so any preferential alignment near the wall arises solely from the active flows generated within the system. The distributions observed in Fig.~\ref{fig:distributions}(a),(b) are consistent with two established results in the literature. First, the planar alignment of extensile systems and homeotropic alignment of contractile systems are the same as those reported for the active anchoring induced by flow fields at nematic–isotropic interfaces,\cite{blow2014biphasic, coelho2021director} and on rigid walls in mixtures of active nematic and passive isotropic fluids.\cite{bhattacharyya2025active} Second, individual microswimmers are known to orient near walls due to hydrodynamic interactions, with pushers aligning parallel and pullers aligning perpendicular to the boundary,\cite{berke2008hydrodynamic, zottl2014hydrodynamics} again consistent with the trends observed here. The agreement across these distinct examples highlights the robustness of flow-induced mechanisms in generating boundary alignment of active particles.

\begin{figure}[h]
\centering
  \includegraphics[height=4cm]{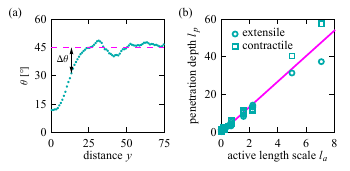}
  \caption{(a) Median angle between director and wall as a function of distance from the wall for $\xi=0$ at $\zeta=0.002$ and $K_\textbf{Q}=0.01$. Dashed line shows $45\degree$, corresponding to a fully random director orientation. We define the penetration depth as the distance from the wall at which the difference $\Delta \theta$ between the median angle and $45\degree$ is less than $15\degree$. (b) Anchoring penetration depth as a function of the active length scale $l_a=\sqrt{K_\textbf{Q}/|\zeta|}$. Points are shown for all combinations of $\zeta\in\{0.2,0.02,0.002,0.0002\}$ and $K_\textbf{Q}\in\{0.01, 0.005, 0.001, 0.0005, 0.0001, 0.00005, 0.00001\}$. Magenta line shows fitted function $l_p=\alpha l_a$, where $\alpha=6.75$ is a fitting parameter.}
  \label{fig:penetration}
\end{figure}

Far from the boundaries, active turbulence dominates and the director field evolves continuously in time. We therefore examine how far the wall-induced alignment penetrates into the system. Since the angle between the director and the wall, $\theta$, is restricted to the interval $\left[0\degree,90\degree\right]$, its mean would be  $45\degree$ in the absence of any preferred orientation. We define the penetration depth $l_p$ as the distance from the wall at which the median angle first enters the range $45\degree\pm15\degree$ (Fig.~\ref{fig:penetration}(a)). We find that $l_p$ scales linearly with the active length $l_a=\sqrt{K_\textbf{Q}/|\zeta|}$ (Fig.~\ref{fig:penetration}(b)). This is unsurprising, since correlation lengths in active turbulence typically scale with $l_a$.\cite{hemingway2016correlation}

The alignment at the wall changes for large $|\xi|$ which correspond to the flow-aligning regime. We first focus on the case of $\zeta\times\xi>0$. As the flow aligning parameter $\xi$ increases (decreases) in an extensile (contractile) system, the distribution of the wall alignment angle, $\theta$, broadens (Fig.~\ref{fig:distributions}(c),(d)), indicating a weakening of the preferred planar (homeotropic) orientation. For sufficiently high (low) values of $\xi$, these anchoring states disappear altogether (Fig.~\ref{fig:distributions}(e),(f)). Instead, the probability distribution of $\theta$ develops a maximum at intermediate angles, rather than at $0\degree$ or $90\degree$, for both extensile and contractile systems. The position of this peak varies systematically with the flow aligning parameter $\xi$, revealing the emergence of a distinct flow–alignment–dominated mechanism that governs boundary orientation. Conversely, if $\zeta\times\xi<0$, the system does not choose a single preferred wall anchoring angle at high $|\xi|$, and a broader distribution of $\theta$ is observed at, e.g., $|\xi|=1.5$ (Fig.~\ref{fig:distributions}(g),(h)).

%||||||||||||||||||||||||||||||||||||||||||||||||||||||||||||
%||||||||||||||||||||||FIGURE 2||||||||||||||||||||||||||||||
%||||||||||||||||||||||||||||||||||||||||||||||||||||||||||||

%||||||||||||||||||||||||||||||||||||||||||||||||||||||||||||
%||||||||||||||||||||||||||||||||||||||||||||||||||||||||||||
%||||||||||||||||||||||Analytical theory|||||||||||||||||||||
%||||||||||||||||||||||||||||||||||||||||||||||||||||||||||||
%||||||||||||||||||||||||||||||||||||||||||||||||||||||||||||
\subsection{Analytical theory of flow-induced boundary alignment}\label{subsec:analytics}
To explain the diverse alignment behaviours observed near the walls, we now turn to an analytical description of the underlying mechanisms. While active turbulence persists in the bulk, the flow field is qualitatively modified in the vicinity of rigid boundaries. The no-penetration condition suppresses the velocity component normal to the wall, and the no-slip condition enforces zero velocity at the wall boundary. As a result, the velocity increases with distance from the boundary, producing a flow that resembles simple shear. Under these near-wall conditions, the governing equations simplify, enabling a tractable analytical treatment, as outlined below.

In the immediate vicinity of the wall, the only relevant non-zero component of the velocity gradient tensor is $\frac{\mathrm{d}u_x}{\mathrm{d}y}$, where $u_x$ denotes the velocity component parallel to the wall and $y$ is the coordinate normal to it. Consequently, the balance between viscous and active stresses reduces to
\begin{align}
    2\eta \frac{\mathrm{d}u_x}{\mathrm{d}y} = \zeta Q_{xy},
    \label{eq:stressbalance}
\end{align}
where $Q_{xy} = \frac{3S}{4}\sin{2\theta}$, the $xy$ component of the $\mathbf{Q}$ tensor, $S$ is the magnitude of the nematic order parameter, and $\theta$ denotes the director orientation. Equation~\eqref{eq:stressbalance} assumes that no residual constant stress is present near the wall, similar to the stress-balance arguments used in analyses of spontaneous shear flows.\cite{duclos2018spontaneous}

Substituting $\frac{\mathrm{d}u_x}{\mathrm{d}y}$ into Eq.~\eqref{qdynamics}, extracting the equation for $\theta$, using the standard procedure of Pauli matrices, say,\cite{giomi2012banding} gives an equation for the dynamics of $\theta$,
\begin{equation}\label{eq:theta}
    \dot{\theta}=\frac{3 S \zeta}{8\eta}\left(\frac{3S+4}{9S}\xi\cos\left(2\theta\right)-1\right)\sin\left(2\theta\right).
\end{equation}
We analyse the flow-tumbling and flow-aligning regimes separately.

\medskip
\noindent\textbf{Flow-tumbling regime:} The flow-tumbling regime is defined by~\cite{marenduzzo2007steady}
\begin{equation}
    \frac{3S+4}{9S}|\xi|<1.
\end{equation}
Under this condition, the bracketed term in Eq.~\eqref{eq:theta} is always negative, and $\theta$ admits two equilibrium solutions, $\theta_0=0\degree$ and $\theta_0=90\degree$.

To assess their stability, we set $\theta=\theta_0+\delta\theta$, and linearise for small perturbations $\delta\theta$, yielding
\begin{equation}
    \delta\dot{\theta}=\frac{3 S \zeta \cos\left(2\theta_0\right)}{4\eta}\left(\frac{3S+4}{9S}\xi\cos\left(2\theta_0\right)-1\right)\delta\theta.
    \label{eq:deltathetadot}
\end{equation}
Since the bracketed term remains negative in this regime, stability requires 
\begin{equation}
    \zeta \cos\left(2\theta_0\right) > 0.
\end{equation}
Thus, $\theta_0=0\degree$ (planar anchoring) is stable for extensile systems, whereas $\theta_0=90\degree$ (homeotropic anchoring) is stable for contractile systems, in agreement with the simulation results shown in Fig.~\ref{fig:distributions}(a),(b). However, the uniqueness and stability of these solutions change substantially in the flow-aligning regime in an activity-dependent manner, as discussed below.

\medskip
\noindent\textbf{Flow-aligning regime:} In the flow-aligning regime
\begin{equation}
    \frac{3S+4}{9S}|\xi|>1.
\end{equation}
Under this condition, the bracketed term in Eq.~\eqref{eq:theta} can be either zero or positive giving rise to the following two possibilities:
\begin{itemize}
    \item One of the equilibrium solutions corresponds to the bracketed term in Eq.~\eqref{eq:theta} 
being equal to zero,
\begin{equation}
    \label{eq:leslie}
    \theta_0=\frac{1}{2}\arccos\left[\frac{9S}{\left(3S+4\right)\xi}\right].
\end{equation}
This angle matches the Leslie angle,\cite{edwards2009spontaneous, marenduzzo2007steady} the angle that passive nematogens ($\zeta = 0$) adopt in a simple shear flow. 

To analyse the stability of the director field at the Leslie angle, we linearise around the Leslie angle solution to obtain
\begin{equation}
    \delta\dot{\theta}=-\frac{3S+4 }{12\eta}\zeta \xi\sin^2\left(2\theta_0\right)\delta\theta.
\end{equation}
Therefore, the equilibrium at the Leslie angle is stable if 
\begin{equation}
    \zeta\times\xi>0,
\end{equation}
meaning that flow-aligning extensile (contractile) active nematics at the wall should align at the Leslie angle if $\xi>0$ ($\xi<0)$, in agreement with the simulation results shown in Fig.~\ref{fig:distributions}(e),(f).

\item The second solution arises in the flow-aligning regime because Eq.~\eqref{eq:theta}  admits additional equilibrium orientations $\theta_0=0\degree$ and $\theta_0=90\degree$. Assessing their stability using Eq.~\eqref{eq:deltathetadot} reveals that
\begin{itemize}[label=$\bullet$]
    \item $\theta_0 = 0^{\circ}$ or $\theta_0 = 90^{\circ}$ is unstable if $\xi > 0$ and $\zeta > 0$,
    \item $\theta_0 = 0^{\circ}$ or $\theta_0 = 90^{\circ}$ is stable if $\xi > 0$ and $\zeta < 0$,
    \item $\theta_0 = 0^{\circ}$ or $\theta_0 = 90^{\circ}$ is stable if $\xi < 0$ and $\zeta > 0$,
    \item $\theta_0 = 0^{\circ}$ or $\theta_0 = 90^{\circ}$ is unstable if $\xi < 0$ and $\zeta < 0$.
\end{itemize}
\end{itemize}

%||||||||||||||||||||||||||||||||||||||||||||||||||||||||||||
%||||||||||||||||||||||FIGURE 3||||||||||||||||||||||||||||||
%||||||||||||||||||||||||||||||||||||||||||||||||||||||||||||
\begin{figure}[h]
\centering
  \includegraphics[height=4.5cm]{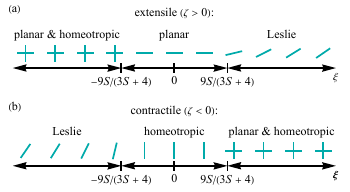}
  \caption{(a,b) Analytical predictions for boundary anchoring. (a) Extensile active nematics exhibit planar anchoring in the flow-tumbling regime. In the flow-aligning regime, they align at the Leslie angle for $\xi>9S/(3S+4)$, while both planar and homeotropic orientations are stable for $\xi<-9S/(3S+4)$. (b) Contractile active nematics exhibit homeotropic anchoring in the flow-tumbling regime. In the flow-aligning regime, they align at the Leslie angle for $\xi<-9S/(3S+4)$, whereas both planar and homeotropic orientations are stable for $\xi>9S/(3S+4)$.}
  \label{fig:prediction}
\end{figure}

The phase behaviour is summarized schematically in Fig.~\ref{fig:prediction}. Extensile systems align parallel to the wall ($\theta=0\degree$) in the flow-tumbling regime. In the flow-aligning regime, they instead adopt the Leslie angle for $\xi>0$, while both planar and homeotropic orientations remain stable for $\xi<0$. Conversely, contractile systems align perpendicular to the wall ($\theta=90\degree$) in the tumbling regime, switch to the Leslie angle for $\xi<0$, and exhibit bistability of planar and homeotropic alignment for $\xi>0$. Note that, in the flow aligning regime, the extensile and contractile cases with the same $\zeta\times\xi$ product are equivalent, reflecting the symmetry of the equations of motion.

\subsection{Comparison to numerical simulations} \label{subsec:comparisons}
\begin{figure}[h]
\centering
  \includegraphics[width=9.2cm]{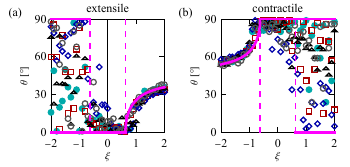}
  \caption{Mode of the angle between the wall and the director at different values of $\xi$ for extensile (a) and contractile (b) active nematics. Full magenta lines show the analytical prediction (for $S=0.35$); dashed lines show the boundary between the flow-tumbling and flow-aligning regions. Values of $(|\zeta|,K_\mathbf{Q})$ used are: ($0.002,0.001$; cyan), ($0.02,0.01$; red), ($0.02,0.0001$; blue), ($0.0002,0.0001$; black), and ($0.0002,0.01$; gray).}
  \label{fig:leslie}
\end{figure}

We now test whether the analytical predictions are borne out in simulations. Figure~\ref{fig:leslie} shows the modal value of $\theta$, the angle between the wall and the director, for extensile and contractile systems across a wide range of the flow aligning parameter $\xi$. The region bounded by the dashed lines in Fig.~\ref{fig:leslie} corresponds to the flow-tumbling regime where the director primarily responds to the rotational component of the shear. We first focus on the extensile case (Fig.~\ref{fig:leslie}(a)), and as expected, extensile systems predominantly adopt the planar configuration, $\theta = 0^{\circ}$, within this interval. 

As $\xi$ increases into the flow-aligning regime, the planar state loses stability and the director reorients to the Leslie angle. The measured angles show excellent agreement with the theoretical prediction of Eq.~\eqref{eq:leslie} (the continuous line) and are independent of both the magnitude of $|\zeta|$ and the elastic constant $K_\textbf{Q}$ (Figs.~\ref{fig:leslie}(a)). Here $\zeta\times\xi$>0 and the active stresses produce flows that reinforce the straining component responsible for alignment, thereby helping to stabilize the Leslie-angle configuration.\cite{nejad2025cellular} 

However, for large negative values of $\xi$, while the analytical calculations suggest two stable configurations - corresponding to both planar and homeotropic orientations - the simulation data is spread across a broad range of orientations (Fig.~\ref{fig:distributions}(g)-(h)) and there is no obvious anchoring. This can be explained by noting that for $\zeta\times\xi<0$, active stresses oppose the aligning strain rate, leading to weak flows and ineffective shear-alignment which is easily disrupted by the vorticity. Analogous behaviour for the different flow aligning parameter regimes is seen for contractile systems (Fig.~\ref{fig:leslie}(b)).

%||||||||||||||||||||||||||||||||||||||||||||||||||||||||||||
%||||||||||||||||||||||FIGURE 4||||||||||||||||||||||||||||||
%||||||||||||||||||||||||||||||||||||||||||||||||||||||||||||

%||||||||||||||||||||||||||||||||||||||||||||||||||||||||||||
%||||||||||||||||||||||||||||||||||||||||||||||||||||||||||||
%||||||||||||||||||||||||||||||||||||||||||||||||||||||||||||
%||||||||||||||||||||||CONCLUSIONS|||||||||||||||||||||||||||
%||||||||||||||||||||||||||||||||||||||||||||||||||||||||||||
%||||||||||||||||||||||||||||||||||||||||||||||||||||||||||||
%||||||||||||||||||||||||||||||||||||||||||||||||||||||||||||
\section{Conclusions}\label{sec:conclusions}

Although confinement is known to strongly influence active flows, a unified framework explaining anchoring developed through wall-induced flows considering both extensile and contractile activity, and both flow tumbling and aligning systems,  has been lacking. Therefore here we use analytical arguments and simulations to investigate active anchoring:
how active nematics align at rigid no-slip boundaries even in the absence of any imposed energetic boundary alignment. 

In the flow-tumbling regime, extensile systems preferentially align parallel to the wall, whereas contractile systems align perpendicular to it. This alignment is not confined to the immediate vicinity of the surface but penetrates into the bulk over a distance proportional to the active length scale, indicating that the physical mechanisms of active anchoring is inherently related to the structure of active flows.

In the flow-aligning regime a qualitative change occurs. Irrespective of the nature of activity, when the product $\zeta\times\xi$ is positive (corresponding to rod-like extensile or disc-like contractile active particles), the director selects the Leslie angle.  When $\zeta \times\xi$ is negative (disc-like extensile or rod-like contractile active particles), no preferred alignment is selected in the simulations due to weak active flows.

The results demonstrate that rigid boundaries can impose effective alignment in dense active matter purely through hydrodynamic flows generated near the walls. This provides a unified interpretation of anchoring phenomena in confined active systems and suggests that flow-induced boundary alignment might contribute to controlling active matter in confinement without the need to engineer specific surface interactions.  

Our framework may also offer a practical route to help classify biological active systems as extensile or contractile, and as flow-tumbling or flow-aligning, in experimental settings where material parameters are difficult to measure directly. More broadly, the role of active anchoring in shaping well-known behaviours of confined active nematics---including spontaneous-flow transitions \cite{voituriez2005spontaneous}, the emergence of unidirectional, bidirectional, oscillatory, vortex-lattice \cite{hardouin2019reconfigurable}, and turbulent states in channels \cite{wu2017transition}, as well as the creation, dynamics, and annihilation of topological defects near boundaries \cite{hardouin2019reconfigurable, hardouin2022active}---remains an avenue for future investigations.

\section*{Author contributions}
MF and JR performed and analysed the simulations. MF, IH, SPT, and JR developed the analytical arguments. SPT, JMY, and JR designed and supervised the research. All authors wrote the manuscript.

\section*{Data availability}
Simulation data used to create the Figures is available at: [Zenodo DOI will be provided before publication].

\section*{Conflicts of interest}
There are no conflicts to declare.

\section*{Acknowledgements}
MF acknowledges support from the Princeton International Internship Program. IH acknowledges support of a Gould \& Watson Scholarship. SPT acknowledges the Royal Society and the Wolfson Foundation for the Royal Society Wolfson Fellowship award, and the support of the Department of Science and Technology, India via the research grant CRG/2023/000169. JMY and JR acknowledge support from the UK Engineering and Physical Sciences Research Council (Award EP/W023849/1) and ERC Advanced Grant ActBio (funded as UKRI Frontier Research Grant EP/Y033981/1). JR acknowledges financial support from the Slovenian Research and Innovation Agency (research project J1-70062, development funding pillars RSF-0106, and research core funding P1-0055).

%%%END OF MAIN TEXT%%%


\balance

%%%REFERENCES%%%
\providecommand{\noopsort}[1]{}\providecommand{\singleletter}[1]{#1}%
\providecommand*{\mcitethebibliography}{\thebibliography}
\csname @ifundefined\endcsname{endmcitethebibliography}
{\let\endmcitethebibliography\endthebibliography}{}

\bibliographystyle{rsc} %the RSC's .bst file

\end{document}